\documentclass[aps,pre,preprint,groupedaddress]{revtex4-1}

\usepackage{amsmath,amsfonts,amsthm} 
\usepackage{graphicx}
\usepackage{xcolor}

\begin{document}


\title{Sharp energy self-determination of macroscopic quantum bodies in pure states, as a validation of the First Law of Thermodynamics}
\author{V\'{\i}ctor Romero-Roch\'in}
\affiliation{Instituto de F\'isica, Universidad Nacional Aut\'onoma de M\'exico \\
Apartado Postal 20-364, 01000 Cd. M\'exico, Mexico}

\date{\today}

\begin{abstract}

We argue that a very large class of quantum pure states of isolated macroscopic bodies have sharply peaked energy distributions, with their width relative to the average scaling between $\sim N^{-1}$ and $\sim N^{-1/2}$, with $N \gg 1$, the number of atoms conforming the body. Those states are dense superpositions of energy eigenstates within arbitrary finite or infinite energy intervals that decay sufficiently fast.  The sharpness of the energy distribution implies that closed systems in those states are {\it microcanonical} in the sense that only energy eigenstates very near to the mean energy contribute to their thermodynamic evolution. Since thermodynamics accurately describes processes of macroscopic bodies and requires that closed systems have constant energy, our claim is that these pure states are typical of macroscopic systems.  
The main assumption beneath the energy sharpness is that the isolated body can reach thermal equilibrium if left unaltered. We argue that such a self-sharpness of the energy in macroscopic bodies indicates that the First Law of Thermodynamics is statistical in character.\\


\end{abstract}

\maketitle

\section{Introduction}

An isolated quantum macroscopic body, whose atoms or molecules interact via short range interatomic potentials, reaches thermal equilibrium if left unperturbed for a sufficiently long time. \cite{LL} We face this situation, for instance, {\it every single time} that the lid of a (very good) thermos bottle is closed or a gas of alkali atoms is confined by a magneto-optical trap in ultra-high vacuum. \cite{Cornell,Ketterle,Pethick,Seman} A very important observation is that this occurs every single time that the experiment is ``repeated'', independently of whether the initial state is the same or not and independently of whether the system reaches the same equilibrium state or not. As a matter of fact, it is actually very hard to conceive or prepare an everyday system, being a solid, fluid or superfluid, to remain in non-equilibrium states, as all tend to thermalize, either in contact with their environment or completely isolated. Our interest here is on isolated systems that do thermalize. In those cases, thermodynamics provides a very precise description of properties and transformations of macroscopic systems in and between states in thermal equilibrium, and gives us very general conditions and restrictions on the relaxation to equilibrium. In particular, the First Law, that establishes that the change of energy of a system during a process equals the change of energy of its surroundings in terms of work and heat, demands that the total energy of the system plus environment, a composite isolated system, remains constant throughout. If we appeal to classical mechanics we can invoke the conservation of energy of closed systems and establish the microcanonical ensemble as the representation of the equilibrium state, in which all points in phase space with energy equal to the initial and constant energy of the system are equally probable. \cite{LL} But, in real life, systems obey quantum mechanics and, in general and strictly speaking, energy is undetermined since quantum states are superpositions of energy eigenstates. To cope with this complication it is usually argued that the energy of a quantum closed system can be determined within a very small ``microcanonical'' shell $\delta E$ around a value $E$ and, then, once equilibrium is reached, the quantum microcanonical distribution is obeyed; see e.g. Refs.  \cite{LL,Goldstein0,Reimann1,Goldstein3}. In particular, Ref. \cite{Goldstein3} provides a thorough discussion of the {\it individual} evolution of these microcanonical states towards equilibrium.\\

However, how can be true, or we be sure, that every time that we prepare a closed system in an arbitrary state it is guaranteed that its energy is within a very small shell $\delta E$? In other words, how can typical quantum states of macroscopic bodies yield a very sharply peaked energy distribution? This is the question that we address here. By ``typical'' states we mean those that we prepare in our every day life or in controlled experiments, as we take for granted that the laws of thermodynamics apply to them. By its enormous encompassing nature, the posed question cannot be answered neither fully nor rigorously. Here, we give very general requirements that pure states $|\psi\rangle$ of macroscopic bodies should obey to obtain such a ``typicallity'' and, for contrast, we also give exceptions to this rule.\\

 As we will discuss, the main requirement that pure states of macroscopic bodies should obey is that they are dense superpositions of energy eigenstates, either {\it bounded} in energy or with a fast decay for large energy. We will qualify what we mean by ``fast'' decay. This demand, in addition to the opposite fact that the  energy density of states of macroscopic bodies that can relax to equilibrium grows extremely fast with energy, allow us to show that very sharply peaked energy distributions are obtained. That is, that the energy of the system is automatically determined within a very small interval $\Delta E$. Hence, the microcanonical shell $\delta E$ around the mean value $E$ is given by $\Delta E$ and does not need to be assumed a priori, as all eigenstates that effectively contribute to the thermodynamic properties have essentially the same constant energy $E$. On the one hand, this allows us to suggest that those states are typical of macroscopic bodies since they lead to the conditions required by the First Law of thermodynamics; and on the other, however, this also implies that the First Law is of statistical character, rather than being an exact or a rigorous one.\\

The result of this paper should also be relevant for the ongoing discussion on thermal equilibration of {\it isolated} macroscopic quantum systems. While this is an old and unabated question, see Refs. \cite{Boltzmann,vanHove,Redfield,Montroll,Zwanzig,Davis,Linblad,Legget,vKOpp,RR,Deutsch,Srednicki,Tasaki,Zurek} to mention a few, there has been a recent vigorous revival of this debate \cite{Goldstein3,Goldstein0,Reimann1,Rigol1,Linden,Deutsch2,Goldstein1,Reimann2,Deutsch3,Kim,Goldstein2,Gogolin,Kaufman,Neill,Eisert,Calabrese,Govinda}, further motivated by recent experimental developments \cite{Kaufman,Neill} 
in which the control on preparation and measurement has yielded powerful tools to test fundamental aspects and assumptions regarding the foundations of statistical physics. In particular, the present study should contribute to the understanding and extensions of the validity of the Eigenstate Thermalization Hypothesis \cite{Deutsch,Srednicki,Rigol1,Reimann2,Kaufman}.\\

In Section II we first revise the well-known fact that the density of states of systems that relax to equilibrium grows extremely fast with energy. \cite{LL} Then, in Section III, we give the general arguments and assumptions on the considered many body pure states $|\psi\rangle$ that yield a sharply peaked energy distribution, and analyze some specific examples.  In Section IV we discuss some exceptions to the rule. Finally, in Section V we briefly review the consistency between equilibrium states and arbitrary but typical pure states, and retake the suggestion that the First Law of Thermodynamics is of statistical character.

\section{Density of states of thermalizing macroscopic bodies and the statement of sharp energy distributions}

Although we have insisted that we consider closed systems that reach equilibrium, as the example of the thermos bottle or the ultracold gases confined in optical or magnetic traps, this is not really a limitation since both thermodynamics and statistical physics assume that composite systems, of an ``open'' system plus environment are, in fact, closed or isolated. \cite{LL} Thus, barring very peculiar cases, such as plasmas or other carefully tailored systems, essentially all bodies that surround us do reach thermal equilibrium if left unaltered. For the sake of argument, consider a chemically pure system in thermal equilibrium. Then, a fundamental thermodynamic result is that any macroscopic {\it subsystem} with a number $N \gg 1$ of atoms or molecules, being part of the closed system in a thermal equilibrium state, will have a single valued, concave, entropy function $S(E,V,N) = N s(e,v)$, with $E$ and $V$ the mean energy and volume of the subsystem, and $s$, $e$ and $v$, their corresponding entropy, energy and volume per particle. By considering systems with positive temperatures only, $s$ is a monotonously increasing function of $e$. Now, as a consequence of the fundamental identification of the entropy in terms of the available number of states $\Delta \Gamma$ of the subsystem, with energy within a very small interval $\delta E$ around $E$, one finds that, quite accurately, \cite{LL} 
\begin{equation}
\Delta \Gamma(E) \simeq e^{N s(e,v)/k_B} \>,\label{denstates}
\end{equation}
with $k_B$ Boltzmann constant. Therefore, for macroscopic subsystems $N \gg 1$, the number of states $\Delta \Gamma(E)$ is an extremely dense function of $E$. However, since the entropy is a concave, monotonously increasing function of the energy, \cite{LL,Callen,MM} we observe that $\Delta \Gamma(E)$ is not only very dense, but also, it is an extremely fast growing function of the energy $E$, for a fixed number of particles $N \gg 1$. A trivial example is the dilute ideal gas, yielding 
\begin{equation}
\Delta \Gamma(E) \simeq C E^{\frac{3}{2}N} \>,\label{ideal}
\end{equation}
with $C$ independent of $E$. And a similar fast growth can be found for Fermi and Bose gases. \\

We now recall that the energy density of states $\omega(E)$ is formally given by 
\begin{equation}
\omega(E) = \frac{d\Gamma(E)}{dE} \>,\label{omega}
\end{equation}
with $\Gamma(E)$ the number of energy states with energy less or equal than $E$. But because the energy spectrum is dense, in any very small energy interval $\delta E$, we can further approximate the density of states as
\begin{equation}
\frac{d\Gamma(E)}{dE} \approx \frac{\Delta \Gamma(E,V,N)}{\delta E} \>, \label{WE20}
\end{equation}
hence indicating that the density of states is also a dense, fast growing function of $E$. We point out now a very important observation: while this result is achieved via the assumptions of statistical physics, this is actually a property of the density of states of the {\it isolated} subsystem of $N$ atoms. That is, it is a property of the Hamiltonian $H$ of the isolated macroscopic subsystem, regardless of the actual state in which it finds itself. This, we will show, is essential for the determination of the energy distribution of macroscopic bodies that can achieve thermal equilibrium.\\
 
Our interest, certainly, is not in the equilibrium states of closed systems but, rather, on arbitrary but typical initial states $|\psi \rangle$ that relax to equilibrium. Hence, we can express such a state as a superposition of energy states of the macroscopic system whose Hamiltonian is $H$,
\begin{equation}
|\Psi \rangle = {\sum_{\{m\}}} a_{\{m\}} |{\{m\}}\rangle \>, \label{Psi}
\end{equation}
with the expansion coefficients,
\begin{equation}
a_{\{m\}} = \langle {\{m\}} | \Psi \rangle \>,
\end{equation}
and where $|{\{m\}}\rangle$ denote the (complete set of) energy eigenstates of the system, $H |{\{m\}}\rangle = E_{\{m\}} |{\{m\}}\rangle$.  We recall here that $|a_{\{m\}}|^2$ is the probability to find the system in the energy eigenstate $|{\{m\}}\rangle$, given that the state of the system is  $|\Psi\rangle$. A different question is to enquire about the probability of finding the system with a value of the energy $E$, say, between $E$ and $E + dE$. Hence, for this matter, let us 
suppose for the moment that we want to know the statistical properties of an operator that involves the Hamiltonian of the system only, $f = f(H)$, when the system is in the state $|\Psi\rangle$ given by Eq. (\ref{Psi}). That is, we want to know the moments of $f$,
\begin{equation}
\langle f^n \rangle = \langle \Psi | f^n(H) | \Psi \rangle \>,
\end{equation}
with $n = 1, 2, 3, \dots$ . These are,
\begin{equation}
\langle f^n \rangle = \sum_{\{m\}} | a_{\{m\}}|^2 f^n(E_{\{m\}}) \>.
\end{equation}
Because the energy levels are dense for the macroscopic system under consideration, we can also write,
\begin{equation}
\langle f^n \rangle = \int  \>  f^n(E) \>{\cal W}(E) dE \>,
\end{equation}
with ${\cal W}(E) dE$ the probability of finding the system with energy between $E$ and $E + dE$. Thus, the statistical properties of $f(H)$ are essentially given by ${\cal W}(E)$.\\

The probability distribution ${\cal W}(E)$ depends on the values of the amplitudes $a_{\{m\}}$, as we amply discuss in the following section, and which, in principle, are quite arbitrary, except that obey
\begin{equation}
{\sum_{\{m\}}} |a_{\{m\}}|^2 = 1\> .\label{norma}
\end{equation}

Two very important quantities associated directly to the distribution ${\cal W}(E)$ are the average energy $\overline E$ and its width $\Delta E$, defined as,
\begin{eqnarray}
\overline{E} &=& \langle \Psi | H | \Psi \rangle \nonumber \\
& = & \int \>   E \>{\cal W}(E) dE  \>.
\label{Emean}
\end{eqnarray}
and 
\begin{eqnarray}
\Delta E^2 &=& \langle \Psi | H^2 | \Psi \rangle - \langle \Psi | H | \Psi \rangle^2 \nonumber \\
& = & \int \>  (E - \overline E)^2  \>{\cal W}(E) dE  \>.
\label{DelE}
\end{eqnarray}

The purpose of this work is to show that there exists a wide class of states $|\psi\rangle$, or equivalently of sets of probability amplitudes $a_{\{m\}}$, such that the ratio of the width $\Delta E$ to the average $\overline E$ of  the distribution ${\cal W}(E)$, scales as 
\begin{equation}
\frac{\Delta E}{\overline E} \sim N^{-\kappa} \label{scales}
\end{equation}
with $1/2 \le \kappa \le1$. If this is true, then we say that the distribution of energy is sharply peaked, since $N \gg 1$. However, since the distribution of energy does not evolve in time, we can claim that the energy remains ``constant'' within a very small interval of energy $\Delta E$, or, that only states whose energy $E$ is very close to $\overline E$ contribute to the determination of the thermodynamics of the system. In the following section we discuss and show the very general requirements that the
states $|\psi\rangle$ should obey to yield a sharp energy distribution.

 \section{Conditions for typical pure states to yield sharply peaked energy distributions}
 
Considering the state $|\Psi \rangle$, given by Eq. (\ref{Psi}), as the initial state of the closed system, we now require certain plausible and reasonable properties of the expansion coefficients $a_{\{m\}}$. The first assumption is that the superposition of states, given by Eq.(\ref{Psi}), is ``dense'' within an interval $E_{min} \le E \le E_{max}$, where the bounds are completely arbitrary; we shall argue below on this interval. But the point is that  if not {\it all} states within the interval $|E_{max} - E_{min}|$ are included, at least there are finite regions within such an interval that are densely populated, see Fig. 1 for a couple of examples. Although we do not rule out the possibility that the initial state is an energy eigenstate or a superposition of a few of them, we do not consider those cases as they are ``atypical'' in the following sense. On the one hand, due to the very dense density of states it is almost a practical impossibility to prepare a macroscopic body in a single eigenstate $|\{m\}\rangle$: Landau and Lifshitz \cite{LL} argue that it would take a time $\Delta t \sim e^N$, in any units, to prepare a system in such a state. That is, this time scale imposes the constraint that we cannot have an energy resolution below a certain ``practical'' limit $\delta E$. 
On the other hand, if we could prepare a single isolated energy eigenstate every single time we repeat an experiment, we would obtain an infinitely sharp energy distribution, ETH would apply and the problem would be solved. But we insist, this is not typical of arbitrary preparations of initial states that we do know relax to equilibrium.\\

Second, perhaps the apparently most demanding but simplifying assumption, we require that, since the superposition is assumed dense, the coefficients $a_{\{m\}}$ can be considered to be a smooth function of its energy, that is,
\begin{equation}
a_{\{m\}} \approx a(E_{\{m\}}) \>. \label{aE}
\end{equation} 
In agreement with the first assumption, this is equivalent to require that all states within any very small width $\delta E$ are essentially equally probable.
This is not unreasonable, for if a single or a few states within $\delta E$ were much more probable than the others, it would amount to accept the possibility that one can prepare single eigenstates.\\

With the two previous assumptions we can write the energy distribution as,
\begin{equation}
{\cal W}(E) \simeq | a(E)|^2 \frac{d\Gamma(E)}{dE} \>, \label{WE}
\end{equation}
with $d\Gamma/dE$ the energy density of states. But because of the result given by Eq. (\ref{WE20}) we can further approximate the density of states, yielding
\begin{equation}
{\cal W}(E) \approx | a(E)|^2 \frac{\Delta \Gamma(E,V,N)}{\delta E} \>, \label{WE2}
\end{equation}
with $\Delta \Gamma(E,V,N)$ given by Eq. (\ref{denstates}). As already mentioned, the growth of $\Delta \Gamma(E,V,N)$ as a function of increasing $E$ is guaranteed by the fact that the entropy function $S/N = s(e,v)$
is a concave, monotonic increasing function of $E$, \cite{LL,Callen,MM} at least for systems that can have positive temperatures only; we do not consider the possibility of negative temperatures here. \cite{Ramsey,Bloch,VRR2,Frenkel} \\

\begin{figure}[h!]
\centering
\includegraphics[width=0.9\linewidth]{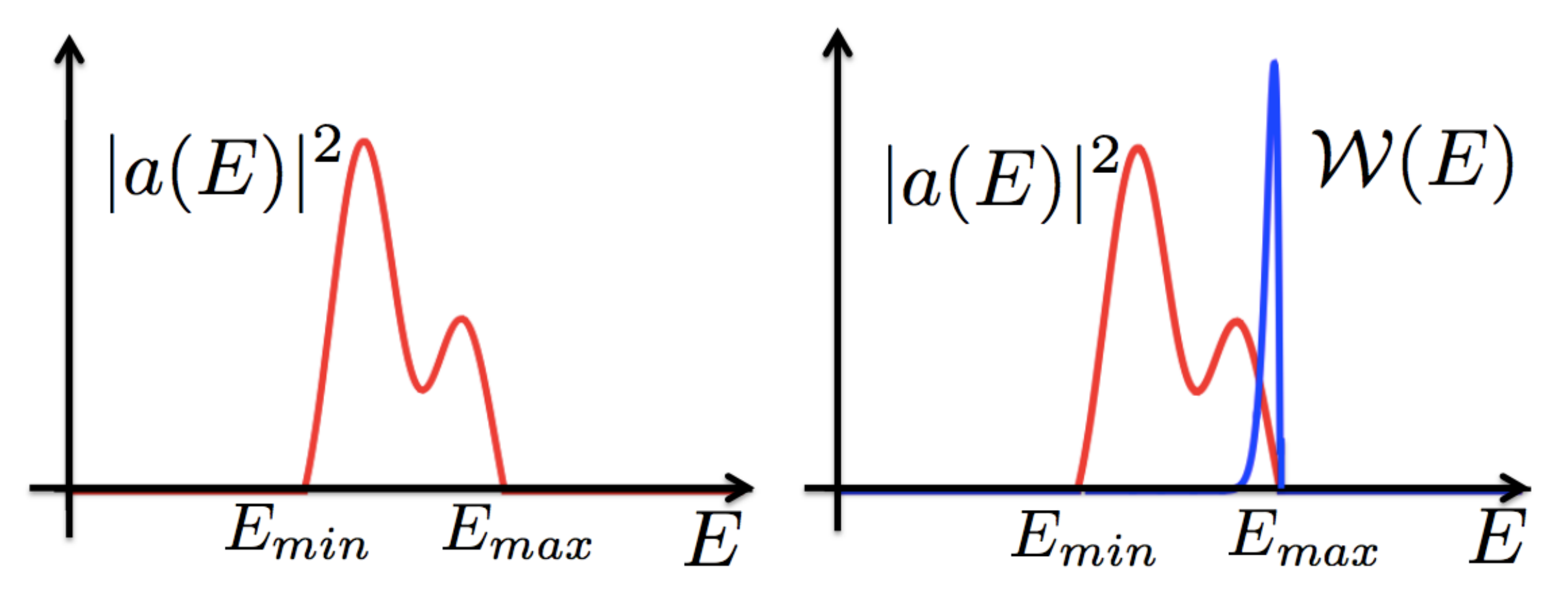}
\\
\includegraphics[width=0.9\linewidth]{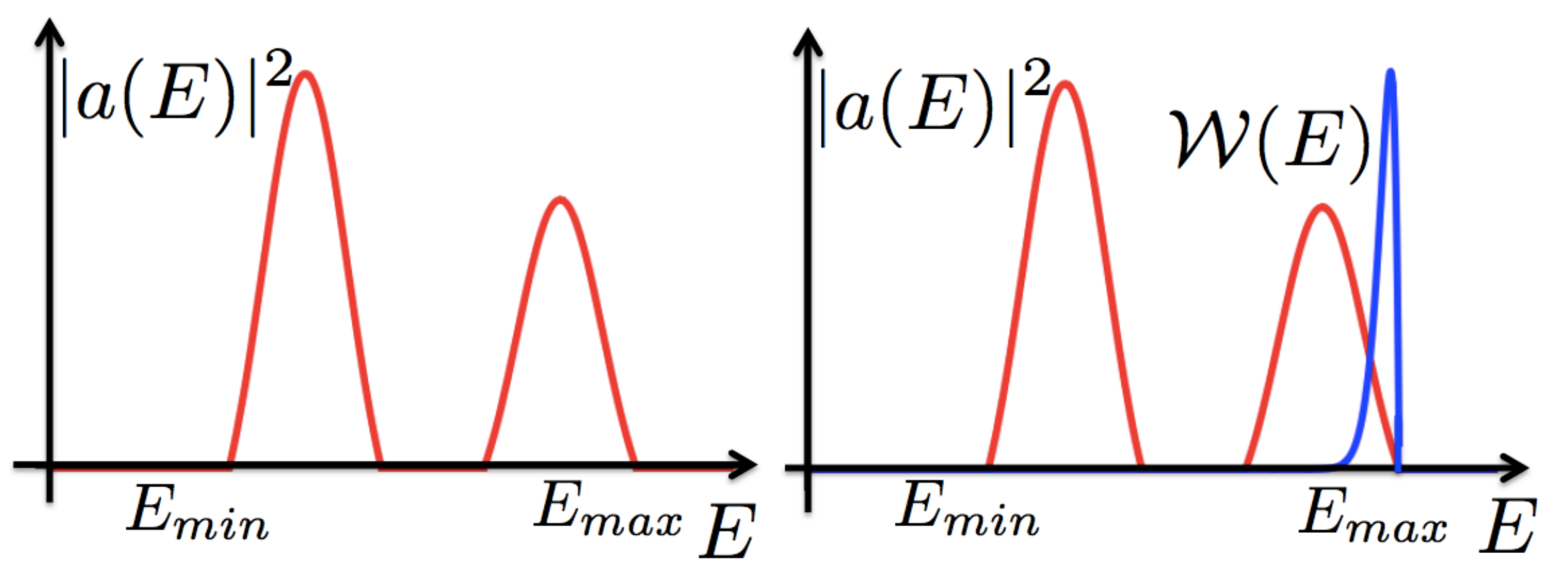}
\caption{(Color online) Sketches of expansion coefficients $|a(E)|^2$ (red line, not at scale) and energy distribution ${\cal W}(E)$ (blue line, not at scale), Eq. (\ref{WE2}), as functions of energy $E$, for two arbitrary initial states $|\Psi\rangle$. For the calculation of ${\cal W}(E)$ we used the density of states of the dilute ideal gas, given in Eq. (\ref{ideal}). }
\label{Figs}
\end{figure}

 We can now examine the assumption concerning the statement that the interval $(E_{min},E_{max})$ is {\it bounded} or decreasing very fast as $E \to E_{\max} \to \infty$. First, let us analyze the ``strict''  case where  $|a(E)|^2 = 0$ for $E$ outside the given energy interval, with $E_{max} < \infty$, as exemplyfied in Fig. 1; that is, when the energy interval is truly bounded. We discuss further below that this assumption can be relaxed. Its purpose is to argue that such a condition immediately implies that  ${\cal W}(E)$ is sharply peaked at an average value $\overline E$, which is smaller but very near $E_{max}$. This can be seen from the fact that, see Eq. (\ref{WE2}), ${\cal W}(E)$ is the product of the density of states that grows very fast without bound and the coefficients that are smooth and bounded, $|a(E)|^2 < 1$, in the energy interval. As a result, the product $|a(E)|^2 \Delta \Gamma(E)$ will necessarily accumulate at the highest possible value of $|a(E)|^2$, yielding a peaked function, as illustrated in the right panels in Fig. 1. As a matter of fact, the value of 
 $E_{min}$ appears to be irrelevant; further, one could even have separated finite regions where $|a(E)|^2 \ne 0$ and the distribution ${\cal W}(E)$ would still be peaked near $E_{max}$. This seemingly simple result indicates that, as long as the probability of occurrence of the energy states is dense and bounded, the energy distribution will be peaked near its highest energy value. In the remaining of this section we justify more explicitly this result, using reasonable assumptions regarding the function $|a(E)|^2$.\\
 
First,  we can estimate how far $\overline E$ is from $E_{max}$ and how sharp is the decay, if strictly $|a(E)|^2 = 0$ for $E \ge E_{max}$. To this end, using Eq. (\ref{denstates}), we write the energy distribution, Eq. (\ref{WE2}), as
\begin{equation}
{\cal W}(E) \approx \frac{1}{\delta E} \exp \left[ \ln |a(E)|^2 + N s(e,v)/k_B\right] \>. \label{WEe}
\end{equation}
Since the exponential function is a monotonic function, the maximum of ${\cal W}(E)$ also occurs at the maximum of its argument. Call $\overline E$ the value of the energy at the maximum. This maximum is determined by the condition that the first derivative of the exponent in Eq. (\ref{WEe}) vanishes, giving the following condition,
\begin{equation}
\frac{1}{k_B}\left(\frac{\partial s}{\partial e}\right)_{\overline E} + \frac{1}{|a(\overline E)|^2}\left.\frac{d |a(E)|^2}{dE}\right|_{\overline E} = 0 \>.\label{barE}
\end{equation}
The first term is positive and the second one negative. The first one equals the inverse of the temperature function of the system at the value $\overline e = \overline E/N$
\begin{equation}
\frac{1}{k_B}\left(\frac{\partial s}{\partial e}\right)_{\overline E} = \frac{1}{k_B T(\overline e,v)} \label{temp}
\end{equation}
and, although initially this is not the temperature of the system, it will become so when the system reaches thermal equilibrium.
The second term in Eq. (\ref{barE}) is negative because of the assumption on the shape of $|a(E)|^2$. Let us assume first a simple but very general model of how $|a(E)|^2$ vanishes as $E \to E_{max}^-$. In this case,  $E$ vanishes algebraically when approaching $E_{max}$,
\begin{equation}
|a(E)|^2 \sim K \left\{
\begin{array}{ccc}
(E_{max} - E_0)^\alpha - (E - E_0)^\alpha &{\rm if} & E \le E_{max} \\
0 &{\rm if} & E \ge E_{max}
\end{array}\right. \label{modela}
\end{equation}
where $\alpha > 0$, $E_0$ is an energy smaller than $E_{max}$, and $K$ is a constant of proportionality. Let us now argue about the properties of common energies of macroscopic systems. To begin with, these energies can be considered to scale with $N$ in the sense that $E/N$ is an energy of the order of $k_B T_{eff}$, with $T_{eff}$ an ``effective'' temperature that may range from fractions of Kelvin, $10^{-7}$ K, at the ground state of  a weakly interacting Bose gas \cite{Bose}, to, say, $10^{10}$ K, at the core of a supernova \cite{supernova}. This is approximately a range from $10^{-30}$ to $10^{-13}$ Joules. We call ``macroscopic'' energies to values of $E/N$ in such an interval, and we expect all energies involved, $E_{max}$, $E_0$ and $\overline E$ to be within it. This allows us to say that those energies scale with $N$, namely, that they are extensive in that sense. \\

Now, because $\overline E$ is very near $E_{max}$, we can write $\overline E = E_{max} - \epsilon$ in Eq. (\ref{barE}), using Eq. (\ref{modela}); since the first term in Eq. (\ref{barE}) is slowly varying in energy, it can be evaluated at $\overline E \approx E_{max}$, while the second one can be expressed in terms of $\epsilon$. The result is that
\begin{equation}
\epsilon \approx \frac{k_B}{\left(\frac{\partial s}{\partial e}\right)_{E_{max}}}
\end{equation}
namely, $\epsilon \approx k_B T(e_{max},v)$ with $e_{max} = E_{max}/N$. In the light of the discussion of the previous paragraph, $\epsilon$ is an intensive quantity, being of order ${\cal O}(1)$ with respect  to $E_{max} \sim {\cal O}(N)$; a very small shift. Now, we can estimate the width of the distribution ${\cal W}(E)$. For this, we consider the second order term in the energy expansion of the argument of the exponential in Eq. (\ref{WEe}). Using the assumed form of the coefficients, Eq. (\ref{modela}),  this yields,
\begin{eqnarray}
&\frac{1}{2}\left(\frac{1}{k_B N}\left(\frac{\partial^2 s}{\partial e^2}\right)_{\overline E} -  \frac{1}{|a(\overline E)|^4}\left(\frac{d |a(E)|^2}{dE}\right)_{\overline E}^2 +  \frac{1}{|a(\overline E)|^2}\left.\frac{d^2 |a(E)|^2}{dE^2}\right|_{\overline E}\right) (E - \overline E)^2 \approx& \nonumber\\
& \frac{1}{2}\left(\frac{1}{k_B N}\left(\frac{\partial^2 s}{\partial e^2}\right)_{\overline E} -  \frac{1}{\epsilon^2} -  \frac{\alpha-1}{\epsilon(E_{max}-E_0)}\right) (E - \overline E)^2 \approx & \nonumber \\
& -\frac{1}{2\epsilon^2} (E - \overline E)^2 .
\end{eqnarray}
The last approximation follows from the fact that the first and third terms in the second line scale as $1/N$, and the second one as ${\cal O}(1)$. Therefore, the energy distribution approximates as,
\begin{equation}
{\cal W}(E) \approx |a(\overline E)|^2\frac{e^{Ns(\overline e,v)/k_B}}{\delta E} \exp \left[- \frac{(E - \overline E)^2}{2\epsilon^2}\right] \label{Wgauss}
\end{equation}
with $\overline E \approx E_{max} - \epsilon$ and $\Delta E = \epsilon \approx k_B T(\overline e,v)$. The above distribution is an extremely sharp gaussian function, since $\overline E \sim {\cal O}(N)$ and $\Delta E = \epsilon \sim {\cal O}(1)$; namely, $\Delta E/\overline E \sim N^{-1}$. It is much narrower than the usual thermodynamic gaussians, whose widths scale as $\sim N^{1/2}$. \\

While the previous argument was done for $|a(E)|^2$ approaching zero as $E \to E_{max}^-$ algebraically, the same can be shown to obtain if it does so exponentially, namely,
\begin{equation}
|a(E)|^2 \sim K \left\{
\begin{array}{ccc}
e^{-\left((E-E_0)/E_1\right)^\gamma} - e^{-\left((E_{max}-E_0)/E_1\right)^\gamma} &{\rm if} & E \le E_{max} \\
0 &{\rm if} & E \ge E_{max}
\end{array}\right. , \label{modela2}
\end{equation}
with $\gamma > 0$, $K$, $E_0$ and $E_1$ constants, but with  both $E_0 \sim {\cal O}(N)$ and $E_1 \sim {\cal O}(N)$, such that the interval $|E_{max} - E_{min}|$ is macroscopic. The ensuing distribution ${\cal W}(E)$ again behaves as given by Eq. (\ref{Wgauss}). 
There are, however, other possible behaviors of the coefficients $|a(E)|^2$ that we now address.\\

In the above paragraphs we discussed cases in which $|a(E)|^2 = 0$ for $E \ge E_{max}$. However, this requirement can be relaxed and demand now that $|a(E)|^2 \to 0$ exponentially, as $E \to \infty$. That is, let us assume that for large $E$, $|a(E)|^2$ behaves as,
\begin{equation}
|a(E)|^2 \sim  \exp[-(E/\Delta)^\kappa]  , \>\>\>{\rm for}\>\>\> E \to \infty \label{expo} 
\end{equation}
with $\Delta$ a scale of energy whose value we discuss below. If $\kappa \ge 1$, because $Ns(e,v)$ is concave, there exists always a maximum in ${\cal W}(E)$, as can be seen by writing the energy distribution for large $E$, see Eq. (\ref{WEe}),
\begin{equation}
{\cal W}(E) \approx \frac{1}{\delta E} \exp \left[ -(E/\Delta)^\kappa + N s(e,v)/k_B\right] \>\>\>{\rm for}\>\>\> E \to \infty. \label{WE3}
\end{equation}
For $0 < \kappa < 1$ the maximum may not exist. Let us discuss first $\kappa \ge 1$. The maximum value $\overline E$ is found from Eq. (\ref{WEe}),
\begin{equation}
\frac{1}{k_B}\left(\frac{\partial s}{\partial e}\right)_{\overline E} \approx  \kappa \frac{\overline E^{\kappa -1}}{\Delta^\kappa} ,
\label{barE2}
\end{equation}
while the width of the distribution $\Delta E$ may be obtained from the second order expansion term, already identifying the width,
\begin{eqnarray}
-\frac{1}{2\Delta E^2}(E -\overline E)^2 & = & \frac{1}{2} \left[ \frac{1}{Nk} \left(\frac{\partial s^2}{\partial e^2} \right)_{\overline E} - \kappa(\kappa -1) \frac{\overline E^{\kappa -2}}{\Delta^\kappa} \right] (E -\overline E)^2 \nonumber \\
&=& \frac{1}{2} \left[ \frac{1}{N k} \left(\frac{\partial s^2}{\partial e^2}\right)_{\overline E} - (\kappa -1) \frac{1}{\overline E k_B}\left(\frac{\partial s}{\partial e}\right)_{\overline E} \right] (E -\overline E)^2 \label{DelE2} .
\end{eqnarray}
First, both terms within the square brackets are negative, the first one because the function entropy is concave and the second because of the assumption $\kappa \ge 1$. Now it comes an interesting point. If we stick to the requirement that $\overline E \sim {\cal O}(N)$, such that the entropy $s = s(e,v)$ is intensive, then both terms in the square brackets are of the same order, yielding a width $\Delta E \sim {\cal O}(N^{1/2})$. This case, however, demands a further requirement on $\Delta$ as seen from Eq. (\ref{barE2}); that is, it should be true that $\overline E^{\kappa -1}/\Delta^\kappa \sim {\cal O}(1)$ for the derivative of $s$ with respect to $e$ to be intensive, yielding $\Delta \sim {\cal O}(N^{(\kappa -1)/\kappa})$. Thus, as long as $\kappa \ge 1$ and we demand that the function $s(e,v)$ and $e$ are intensive always, then, the exponentially decaying function $a(E)$, as given by Eq. (\ref{expo}), also gives rise to a sharply peaked energy distribution with width $\Delta E \sim {\cal O}(N^{1/2})$. For $\kappa = 1$ the reader can see that this is the usual textbook argument to show that the canonical equilibrium distribution yields sharply peaked energy distributions, if $\Delta$ is the temperature of the system in equilibrium. \cite{LL}\\

However, what if $\Delta$ in Eq.(\ref{expo}) does not scale in the way described in the previous paragraph? For instance $\Delta$ could be $\sim {\cal O}(1)$ or  $\sim {\cal O}(N)$. Although we cannot strictly make general statements for an arbitrary system, we can check different cases with the dilute ideal gas, given in Eq. (\ref{ideal}), as we can calculate explicitly the expressions in Eqs. (\ref{barE2}) and (\ref{DelE2}). We find that $\overline E$ does not scale with $N$, in general; for instance, if  $\Delta \sim {\cal O}(1)$, $\overline E \sim {\cal O}(N^{1/\kappa})$, while if $\Delta \sim {\cal O}(N)$, $\overline E \sim {\cal O}(N^{1+ 1/\kappa})$. In the same fashion, $\Delta E$ scales differently for each case. While these kinds of scaling of $\overline E$ with $N$ seem to be unusual, we expect nevertheless that $\overline E$ should be within realistic bounds as described above. The notorious result is that, for all type of dependences, the ratio $\Delta E/\overline E \sim {\cal O}(N^{1/2})$ always. That is, the distribution ${\cal W}(E)$ is always sharply peaked, with a relative width $\sim N^{-1/2}$. \\

\section{Exceptions to the rule}

Clearly, the arguments given above fail if the product of $|a(E)|$ times $\Delta \Gamma(E)$ has a long tail as $E\to \infty$, as this would yield an arbitrarily large width $\Delta E$. But this is only possible if $|a(E)|$ decays much more slowly than the ensured growth of $\Delta \Gamma(E)$. Such a behavior can also be illustrated with the exponential form of the coefficients given by Eq. (\ref{expo}), in the case where $0< \kappa < 1$. In this situation there is no guarantee that ${\cal W}(E)$ has a maximum, neither that it is normalized, because both terms in the exponent in Eq. (\ref{WE3}) are concave. However, we can tailor the value of $\kappa$ such that, still, the maximum exists, but we can also tune it such that the width $\Delta E$ is as {\it large} as we desire, see the second line in Eq. (\ref{DelE2}): note that if we choose $\kappa < 1$ appropriately, the gaussian-like exponent in ${\cal W}(E)$ can still be negative, yet we could make the factor multiplying $(E-\overline E)^2$ arbitrarily small in absolute value. A similar reasoning can be used if we choose that the coefficients $|a(E)|^2$ vanish as $E \to \infty$ algebraically, say $|a(E)|^2 \sim E^{-\eta}$. Choosing $\eta$ appropriately, we can make the tail of ${\cal W}(E)$ to behave as slow as we desire, but still normalizable, and obtain a very large width $\Delta E$. These two cases make the general statement of this article to fail, yet we claim, doing this requires a very detailed tailoring of the initial state $|\Psi \rangle$. That is, the coefficients must be strictly different from zero as $E \to \infty$ and must almost exactly cancel the enormous growth in energy of the density of states, in order to render a slow decay of the energy distribution. This appears just as complicated, perhaps, as trying to obtain a superposition of very few states. In any case, it would certainly be very interesting to be able to prepare in real life states of macroscopic systems with such long energy probability tails, for after measurement, one would obtain a very different energy each time the system were prepared in the {\it same} state.\\

To summarize our claim, we can state that a very sharp energy distribution is obtained for initial pure states $|\Psi\rangle$, whose energy coefficients $|a(E)|^2$ are bounded from above by an energy value $E_{max}$ or decay to zero as $E \to \infty$ sufficiently fast. An important point is that states whose energy is well below $\overline E$ are quite irrelevant. This has an interesting consequence when the superposition of states which is dense but in ``lumps", such as that shown in the lower panel of Fig. 1. In such a situation the interference between states of different lumps would not affect the thermodynamics of the system, because its energy would effectively remain in those states near the maximum value of the energy $E_{max}$ all the time.\\

\section{Final Comments} 

The claim of this paper rests on assuming that the systems under consideration can reach thermal equilibrium. While we have no intention of indicating how this does occur, we still have few questions that should be addressed regarding the consistency or compatibility of an equilibrium state with the fact that the system is always in a pure state. As already cited in the introductory paragraphs, there are recent excellent discussions, see Ref. \cite{Goldstein3} and references therein, on how thermal equilibrium is achieved in quantum macroscopic bodies that are in pure states within a microcanonical shell. The issue we address in this paragraph regards the equilibrium state itself.  As discussed in Landau and Lifshitz monography and in the recent papers, equilibrium is reached when distributions of extensive quantities of {\it macroscopic subsystems} of the whole isolated body are sharply peaked; in this case, it is true that $\Delta A_s/\overline A_s \sim {\cal O}(N_s^{-1/2})$, with $A_s$ an extensive quantity of the $s-$th subsystem and $N_s \gg 1$ its number of particles. This condition is achieved as a consequence of the macroscopic subsystems becoming statistically independent.  Concomitantly, one can assert that states of any subsystem, with the same number of particles, volume and, specially, same energy, are equally probable. This in turn allows us to establish that the equilibrium state of the subsystems can be accurately described by a density matrix $\rho_s$, such that the entropy of the $s-$th subsystem, $S_s$ can be obtained from the expression, \cite{LL}
\begin{equation}
S_s = - k_B {\rm Tr_s}\> \rho_s \ln \rho_s ,\label{entropy}
\end{equation}
where the trace is taken over states of the system $s$ alone. We can now proceed ``backwards'' and conclude that, regarding the whole isolated system, its state of equilibrium can be described by the microcanonical density matrix that asserts that all states within a very narrow energy band are equally probable, and that states outside of it have probability zero. \\

The above well-known assertions regarding equilibrium states may seem contradictory with the fact that we are assuming that the system is always in a pure state $|\Psi\rangle$, with a quite arbitrary but typical superposition of energy eigenstates. First, since the system is isolated, its time evolution is unitary under its Hamiltonian $H$ 
\begin{equation}
|\Psi(t) \rangle = e^{-iHt/\hbar} |\Psi\rangle \>,\label{psit}
\end{equation}
and, therefore, the energy distribution ${\cal W}(E)$ remains stationary for all times. Second, as essentially only energies corresponding to energy states within $\Delta E$ around $\overline E$ are probable, we can then assert that, in the evolution of $|\Psi(t)\rangle$, the energy of the system remains constant within such an interval. And without compromising an explanation of how it occurs, the system reaches equilibrium. But the point is that we can attach to such an equilibrium state the actual microcanonical ensemble, with $\sim e^N$ states in $\Delta E$ around $\overline E$, since all have the ``same'' energy and are ``equally'' probable. It is therefore clear that only when the system has reached equilibrium we can use the microcanonical density matrix as a {\it representation} of the equilibrium state, regardless of whether it is in a pure state, known or unknown. As a clear consequence, all initial ``typical'' states $|\Psi\rangle$ with approximately the same mean energy $\overline E$, once in equilibrium, can be described by the same microcanonical density matrix. After all, this density matrix permits calculation of thermodynamic properties only, including their correlations, which usually are properties of few bodies; of course, those properties can also be calculated using the state $|\Psi(t)\rangle$, if known. In this way, using the microcanonical density matrix in Eq. (\ref{entropy}), leads to the entropy of the equilibrium state as being (the negative of the logarithm of) the number of energy states within $\Delta E$ around $\overline E$, that is, to the expected value $S = - k_B \ln \Delta \Gamma(\overline E,V,N)$. It is certainly erroneous to substitute the state $|\Psi(t)\rangle$ into the formula given in Eq. (\ref{entropy}) since it would yield the absurd result that the entropy is zero for an isolated system in thermal equilibrium.\\

To conclude, we mention once again that the present result indicates that the First Law of Thermodynamics may be of statistical character, just as we are used to the fact that the Second one indeed it is. Usually, in order to enunciate the First Law one appeals to the conservation of energy of isolated classical systems. Then, one argues that the change in energy of a system must equal the corresponding change in energy of its surroundings in terms of heat and work. However, real systems obey quantum mechanics and their energy cannot be considered to be a constant, unless they are in an energy eigenstate. But as this cannot be ensured every single time we prepare a macroscopic system, the message of this paper is that 
macroscopic bodies are naturally prepared in states whose energy distribution are sharply peaked. In turn, this entails us to affirm that the energy of the system {\it effectively} remains constant, and the First Law can then be established. The striking conclusion is that such a law has a statistical validity, not a rigorous nor an exact one. As a matter of fact, since classical mechanics emerges from quantum mechanics for macroscopic systems, one could also claim that the conservation of energy of classical systems is a consequence of the energy sharpness of the corresponding quantum bodies. This opens the possibility of preparing isolated macroscopic systems in atypical initial states, always in the same one, that would yield very distinct values of its energy after measurement, in ``violation'' of the First Law of Thermodynamics. \\

{\bf Acknowledgement.} Acknowledgement is given to grant PAPIIT-IN108620 (UNAM).


\begin{thebibliography}{999}
\bibitem{Bose} The ground state energy, per atom, of a monoatomic weakly interacting Bose gas \cite{Pethick} is given by $E_0/N = (2 \pi \hbar^2 a/m) N/V$, with $a > 0$ the scattering length. Using typical data of recent ultracold gases \cite{Pethick}, say $^{87}$Rb with $a \approx 50 \times 10^{-8}$ cm and $N/V \approx 10^{15}$ atoms/cm$^3$, one obtains $E_0/N \approx 10^{-23}$ erg.
\end{thebibliography}
\end{document}